\documentclass[11pt,twoside]{article}
\usepackage{newpasp}

\usepackage{epsf}

\index{Sancisi, R.}
\index{Fraternali, F.}
\index{Oosterloo, T.}
\index{van Moorsel, G.}

\begin{document}

\title{The HI halos of spiral galaxies}
\author{Renzo Sancisi}
\affil{Astronomical Observatory, Bologna, It \& Kapteyn Inst., Groningen, NL}
\author{Filippo Fraternali}
\affil{I.R.A.(C.N.R.) \& Astronomy Department, Bologna, It}
\author{Tom Oosterloo}
\affil{N.F.R.A., Dwingeloo, NL}
\author{Gustaaf van Moorsel}
\affil{N.R.A.O., Socorro, NM, USA}

\begin{abstract}
{The results of observational studies on the vertical HI density distribution
and kinematics of the disks of spiral galaxies are reported and discussed.
Attention is drawn to the presence of HI emission, unknown before and here
referred to as the `beard', which has anomalous structure and kinematics,
different from the known cold HI disk.  This component is extended and
probably located in the halo region, it is rotating more slowly than the disk and
shows radial inward motion.  Its nature and origin $-$galactic fountain or
infall of primordial gas?$-$ are still a puzzle.}
\end{abstract}


\section{Introduction}

The vertical structure of the HI disks of spiral galaxies has been the subject
of various investigations in recent years.  Warping and flaring seem to be
common features of the outer HI layers and have been, since their discovery,
at the center of the attention mainly because of the key information they may
hold on the dark matter halos.  Van der Kruit (1981) and more recently Olling
(1995) have pioneered the use of the flaring of the HI layers of edge-on
galaxies to determine the mass of the stellar disks and the shape of the dark
matter halos.

Also in the inner parts of the disks there is evidence of vertical extensions
and motions of the HI.  These, however, are probably related to the star
formation activity in the disk and form part of the gas circulation between
disk and halo. Their study is important for understanding galaxy
evolution. For instance, they may reveal systematic gas in- and out-flows from
the disk. They may help understanding the
connection between all the phases of the interstellar medium.  Also, such a
study should provide new insights in the origin and nature of the HVCs in our
galaxy.

Besides these processes internal to the disk there are other events of
external origin, like the interactions with companions and the possible infall
of primordial intergalactic gas, which may affect the density distribution and
kinematics of the gas in the halo region and also play an important role in
the formation and evolution of disks (see for a recent review Sancisi, 1999).

\section{Edge-on and face-on galaxies}

For the study of the disk-halo connection edge-on and face-on systems have
been observed, the former to obtain the projected vertical distribution of the HI
density, and the latter to reveal the vertical motions.  In edge-on galaxies
the separation of an HI thick disk or halo component from the 
warping and flaring of the outer disk is very difficult because of
projection effects and requires a very careful analysis of the system
kinematics.

NGC~891 offers a good illustration of these problems and ambiguities.  The HI
pictures obtained by Rupen (1991) and by Swaters, Sancisi \& Van der Hulst
(1997) show the presence of emission extending up to about 2 arcmin (5.6
kpc) on both sides of the galaxy plane (Fig. 1).  This has been interpreted in
the past (Sancisi \& Allen, 1979) as being part of a huge outer flare of the
gas layer seen in projection.  Also the possibility of a line-of-sight warp
has been discussed (Becquaert \& Combes, 1997).  But Swaters {\it et al.}
(1997) have concluded from their 3D analysis of the HI data cube that the
extended emission comes from gas not in the outer flare but in the inner halo
region of NGC~891 and have proposed that this gas rotates more slowly (25 to
100 km s$^{-1}$) than the gas in the plane.  They have tentatively attributed
this velocity difference to a gradient in the gravitational potential and have
pointed out that this may serve to discriminate between disk and spheroidal
mass models.  The HI in the halo has a mass of about 6$\times$ 10$^8$
M$_{\odot}$.  
This amounts to about 15\% of the total HI mass of NGC~891 and
only 0.3\% of the total mass.
There is little doubt that the origin of this halo gas is related to the star
formation activity in the disk of NGC~891 and to the connected disk-halo circulation.
This is supported by the presence of a thick disk component which is observed in the radio
continuum (Dahlem, Dettmar \& Hummel, 1994) and in H$\alpha$ (Rand, Kulkarni \&
Hester, 1990) and is also indicated by the extended dust structures visible in the 
high-resolution optical WIYN images (Howk \& Savage, 1997).

The motion of the gas in the vertical direction from and toward the galaxy
plane has been studied in a number of face-on galaxies. Well-known examples
are NGC~628 (Kamphuis \& Briggs, 1992), NGC~1058 (Dickey {\it et al.}, 1990)
and NGC~6946 (Kamphuis \& Sancisi, 1993).  These observations have led to
accurate determinations of the gas velocity dispersion and occasionally they
have revealed the presence of gas with velocities of up to 50-100 km s$^{-1}$
perpendicular to the galaxy plane.  In some cases this high-velocity gas is
seen in connection with HI holes in the disk (Kamphuis \& Sancisi, 1993;
Kamphuis, 1993) suggesting a link with star formation activity. Such HI
observations of face-on galaxies do not provide, however, any information on
the z-distribution of the gas.

The limitations and difficulties in the study of the vertical HI density
structure and kinematics that are met with galaxies viewed
either edge-on or face-on can be partly overcome by taking objects of
intermediate inclinations. 
This is illustrated very well by the case of NGC~2403 
which has now become the prototype for such a study.  As it will become
clear below, this nearby spiral galaxy is a most suitable candidate because
of its ideal inclination of about 60$^{\circ}$, its extended HI layer (about
twice the optical) and its regular kinematics and symmetric, flat rotation
curve. And, indeed, the HI data do show all the features seen in edge-on and
face-on galaxies and a more complete interpretation is possible.

\section{Galaxies with `beards'}

The early HI observations with the Westerbork Synthesis Radio Telescope
(WSRT), although it was not recognized at the time, show puzzling systematic
asymmetries in the HI velocity profiles along the major axis of NGC~2403. 
The asymmetries are in the form of wings on the side of the
lower velocities, towards systemic.  This pattern (we refer to it as the
`beard' and show it in Fig. 3 below) is clearly visible 
in the position-velocity maps along the major axis
of NGC~2403 (Wevers, Van der Kruit \& Allen, 1986, p.  551, Fig. 6a; Begeman,
1987, p.  46, Fig. 3). It is similar to that found in edge-on galaxies and
also in objects observed at relatively low angular resolution.  Those
observations of NGC~2403 have, however, a sufficiently high angular resolution
and the galaxy is not too highly inclined (i$\sim$60$^{\circ}$).  It is,
therefore, quite surprising to see such systematic asymmetries in the line
profiles.

Recently Schaap, Sancisi \& Swaters (2000) have re-analysed the 21-cm line
observations with higher angular resolution and better sensitivity which had
been obtained by Sicking (1997) with the WSRT.  In
particular, they have investigated the shape of the HI velocity profiles
and confirmed the presence of the asymmetric structure.  They have also
discussed its nature and origin.  It is clear that it cannot simply be
explained by gas moving perpendicularly away from or toward the disk, because
that would produce extensions symmetric with respect to the rotation velocity.
They have investigated therefore the effects which are produced on the shape
of the HI velocity profiles by the vertical structure of the HI layer.  In
order to do this they have constructed 3D models of the galaxy examining the
possibilities of $i)$ a thick HI disk, as opposed to the usually assumed thin
layer, and $ii)$ a two-component structure, with a thin layer and a thicker
but less dense one.  They have concluded that, in order to reproduce the
observations, the thick HI disk of the one-component model considered in $i)$
would have to be unrealistically thick, at least about 5 kpc at full
half-width.  A better representation is obtained by assuming a thin layer with
gas extensions from the plane which rotate more slowly (about 25-50 km s$^{-1}$)
than the gas in the plane.  Clearly, this is a result similar to that found by
Swaters {\it et al.} (1997) in NGC~891, with gas in the halo region rotating
more slowly than in the plane.

\subsection{The VLA Observations}

We have pursued the investigation of the structure and kinematics of the HI in
the halo of NGC~2403 with new observations with
the VLA in C configuration and 48 hours integration.  These
data have significantly better sensitivity and higher resolution 
than those used by Schaap {\it et
al.} (2000) and have led to a
considerable improvement of the observational picture.  Not only is the
existence of the `beard' confirmed, but new remarkable features have surfaced.
The total HI map and velocity field are shown in Fig. 2.

Fig. 3 shows the position velocity map along the major axis.  Patchy wings of
HI emission are visible extending from the HI ridge, which in the figure is
roughly marked by the rotation curve (dots), towards the systemic velocity on
both sides of the galaxy.  Due to the improved sensitivity, the new data show
that the beard is much more extended than in Schaap {\it et al.} (2000).  The
unexpected, new features are the traces of emission, of very low signal/noise
ratio but unmistakably real, which show up in the second and fourth quadrant
at `forbidden' or `counterrotating' velocities. 

The overall picture that emerges from the new data is that
of an extended zone of gas which seems to `know' about the general pattern of
the disk rotation but has large deviations from it.  We compare this picture
with that expected for a thin, cold HI layer (Fig. 3, right) which follows the
flat rotation curve of NGC~2403 as indicated by the dots.  All the extended
gas seen at velocities significantly different from rotation, whether part of
the extended beard or in the forbidden quadrants of Fig. 3, we call from now
on the `anomalous' gas. Its total mass is about 3$\times$10$^8$ M$_{\odot}$, 10\% of the total HI mass of NGC~2403, 
and $\sim$0.3\% of its total dynamical mass.  These are similar to the
values found for NGC~891.  The spatial structure of this anomalous gas is
visible in a representative sample of channel maps displayed in Fig. 4.  Long
filaments, up to 5-10 kpc, and spurs appear to dominate the HI distribution.
The masses of these structures are between 10$^6$ and 10$^7$ M$_{\odot}$.

Are there any faint tails of emission on the high velocity sides in the
position-velocity map of Fig. 3? On the approaching side it is difficult
to be sure because of the confusion with the galactic HI emission. 
On the receding side, instead, around heliocentric velocities of
270-280 km s$^{-1}$, there are hints of emission at about 25-30 km s$^{-1}$ 
with respect to the rotational velocity. These may represent vertical 
motions, which deprojected would be around 50-60 km s$^{-1}$, and may be due 
to gas actually being ejected from the disk and moving into the halo.

\section{Discussion and Conclusions}

A 3-dimensional model analysis has been carried out and the possibilities of a
single thick gas layer, of an outer flare and of a line-of-sight warp have been
investigated and excluded.  The effect of an extended, low density halo
component has been finally studied.  A two-component structure $-$a thin cold
disk and a thicker layer with separate kinematics$-$ has been assumed.
Some of the results are reported in the poster paper of
Fraternali {\it et al.} (this conference). The first step in the
analysis has been a tentative separation of the anomalous gas component from
the dense and regularly rotating gas layer.  The result is a density map
(Fig. 2, bottom right) and a velocity field (see poster) for the anomalous gas.  Its
kinematics is clearly dominated by differential rotation, but the mean
rotation velocities are more than 50 km s$^{-1}$ lower than in the disk in the
inner parts and about 20 km s$^{-1}$ in the outer parts.  The kinematical
minor and major axes appear to be somewhat rotated as compared to those of the
thin disk and not orthogonal.  This suggests a radial inward motion, probably
increasing in the central region. Detailed 3D modelling confirms this.  

What is the nature and the origin of the anomalous gas?  No 
satisfactory explanation has been found yet for its peculiar morphology and
kinematics. The overall pattern of motion (Fig. 3) and the large-scale
regularity indicate that this gas, despite its anomalous motion, is not
`weird'. Clearly, it represents a basic, integral part of the system. The
`beard' and the apparently `counterrotating' components do not seem
manifestations of two different types of phenomena.  They rather seem to form
together one coherent structure and require, therefore, one unique
satisfactory explanation.  Possibilities are:

1. A galactic fountain (Shapiro \& Field, 1976; Bregman, 1980) powered by
supernova explosions and stellar winds which move the ionized gas from the
disk into the halo. The gas may reach large distances from the plane and then
cool down and fall back onto the plane. 
The hypothesis of such a fountain is
attractive because of the presence of active star formation and bright HII
complexes in the disk of NGC~2403. 
Also the presence of a large number of
holes in the HI layer (see Fig. 2, bottom left) and presumably of supershells
supports the picture of an effervescent disk and of a significant
disk-halo-disk circulation. 
We are further pursuing the study of these
phenomena with deep spectroscopy of the H$\alpha$ emission from the bright inner
parts of NGC~2403.  
In many respects NGC~2403 and NGC~891 seem to be similar.
The main difficulty with a standard fountain interpretation in the case of NGC~2403
lies in the presence of the apparently counterrotating, `forbidden' emission.
In order to explain that, a new approach and different assumptions for the
fountain dynamics, for instance no conservation of angular momentum, may be
necessary.

2. Infall of {\it primordial} intergalactic gas. The hypothesis of cosmological gas
infall has been proposed in the past and discussed (cf. Oort, 1970) in
connection with the problem of the origin of the HVCs of the
Galaxy and with the need to re-supply the disks of spiral galaxies with
fresh gas. Such a possibility has to be considered here.  In that framework it
would be essential to understand the overall kinematical pattern of the
anomalous HI and to understand its large-scale coherent
structure which follows the rotation of the disk of NGC~2403.  Also, it would
be important to determine its metal abundance and to find out whether it is as
low as found for the intergalactic gas or it is closer to solar.

Configurations involving a mis-alignment of
the rotation axes of the halo and of the disk, as proposed in connection with
the origin of warps (Debattista \& Sellwood, 1999), have also been considered
here for NGC~2403. The non-orthogonality of major and minor axis shown by the
velocity field of the anomalous component seems to argue against it and in
favour of radial motions.

Finally, the possibility of a large asymmetric drift of the halo HI 
with a high velocity dispersion and a significant decrease of its rotational
velocity should also be considered.
 
The results of the present HI observations of NGC~2403 have interesting
implications also regarding the High Velocity Clouds observed in the Milky Way
(Wakker \& van Woerden, 1997) which are still a mystery and whose distances,
nature and origin are still a matter of debate.  
Is it possible that the
anomalous HI complexes discovered around NGC~2403 are analogous to the HVCs
seen in the Galaxy? 
With what we know of the HVCs it seems fair to suggest
that we probably have detected in NGC~2403 a population of HI clouds of the
same type as those intermediate and high velocity
clouds which are thought to be closely associated with the Galaxy and to be
probably a galactic fountain type of phenomenon.  Clearly, they would be
different from the Magellanic Stream.

How common are the `beards' of galaxies? As far as we know there are some
cases, like those of M33 and UGC~7766, which
indicate the presence of similar structures. It is clear, however, that HI
observations of higher sensitivity are needed. These
would tell whether there is any correlation with 
luminosity or mass or high surface brightness or
environment, and which of these may play the crucial role.

\end{document}